\documentclass[10pt]{article}
\newcommand {\be}{\begin{equation}}
\newcommand {\ee}{\end{equation}}
\newcommand {\bea}{\begin{eqnarray}}
\newcommand {\eea}{\end{eqnarray}}
\newcommand {\refeq}[1] {(\ref{#1})}

\newcommand {\vett}[1] {\mathbf{#1}}

\newcommand {\pippo}[1]{{\setbox0=\hbox{#1}
  \hbox{\kern-.025em\copy0\kern-\wd0
  \kern.05em\copy0\kern-\wd0
  \kern-0.025em\raise.0433em\box0} }}

\renewcommand{\Im}{\mathrm{Im}}
\usepackage{psfig,graphicx,smallcaptions}
\usepackage{pstricks,pst-coil}

\title{Density profiles and collective excitations
of a trapped two component Fermi vapour}
\author{M. Amoruso$^1$, I. Meccoli$^2$, A. Minguzzi$^1$ and
M. P. Tosi$^{1,3}$\footnote{e-mail: tosim@bib.sns.it} \\
{\it {\small $^1$	Istituto Nazionale di Fisica della Materia and
Classe di Scienze,}}\\   
{\it {\small Scuola Normale Superiore, Piazza dei 	Cavalieri 7,
56126 Pisa, Italy}}\\ 
{\it {\small $^2$ Istituto Nazionale di Fisica della Materia and
Dipartimento di 
Fisica,}}\\ {\it {\small Universit\`a di Parma, Parco Area delle
Scienze 7a, 
43100 Parma, Italy}}\\{\it {\small
$^3$	Abdus Salam International Centre for Theoretical Physics,}}
\\{\it {\small Strada
Costiera 11, 34014 Trieste, Italy}}}

\date{}

\catcode `\@=11
\@addtoreset{equation}{section}
\def\theequation{\arabic{section}.\arabic{equation}}
\catcode `\@=12 

\begin{document}
\maketitle
  
\abstract{
We discuss the ground state and the small-amplitude excitations
of a degenerate vapour of fermionic atoms placed in two hyperfine states
inside a spherical harmonic trap. An equations-of-motion approach is set up
to discuss the hydrodynamic dissipation processes from the interactions
between the two components of the fluid beyond mean-field theory and to
emphasize analogies with spin dynamics and spin diffusion in a homogeneous
Fermi liquid. The conditions for the establishment of a collisional regime
via scattering against cold-atom impurities are analyzed. The equilibrium
density profiles are then calculated for a two-component vapour of $^{40}$K
atoms: they are little modified by the interactions for presently relevant
values of the system parameters, but spatial separation of the two
components will spontaneously arise as the number of atoms in the trap is
increased. The eigenmodes of collective oscillation in both the total
particle number density and the concentration density are evaluated
analytically in the special case of a symmetric two-component vapour in the
collisional regime. The dispersion relation of the surface modes for the
total particle density reduces in this case to that of a one-component
Fermi vapour, whereas the frequencies of all other modes are shifted by the
interactions.}

\vspace{2cm}

{\center PACS. 67.40.Db Quantum statistical theory; ground state, elementary excitations}

\section{Introduction}	
\label{1st}

The experimental realization of Bose-Einstein condensation in
confined vapours of alkali atoms \cite{bec_rb,bec_na,bec_li1,bec_li2}
 has given impulse to the study of
dilute quantal fluids, including vapours of fermionic atoms.
Magneto-optical confinement of fermionic species has been reported for $^6$Li
\cite{bradley95} and $^{40}$K \cite{inguscio}. DeMarco et
al. \cite{demarco} have realized magnetic trapping of $^{40}$K 
atoms in two different hyperfine states corresponding to $|F = 9/2, F_z =
9/2\rangle$ and $|F = 9/2, F_z = 7/2\rangle$, with the possibility of
varying the relative `
concentration of these two components of the vapour up to selective removal
of one of them. Earlier experimental work on double Bose condensates
\cite{myatt,double_cond_dyn_exp_1} 
and some of the related theoretical work on the equilibrium state and on
the excitation properties of bosonic mixtures \cite{dbcond1,dbcond2,dbcond3,dbcond4,bigelow_double_condensate,dbcond5} may also be recalled
at this point.

	The s-wave collisions between pairs of fermions in the same
hyperfine state are suppressed by the Pauli principle, so that to leading
order only p-wave scattering and dipole-dipole magnetic interactions remain
in a one-component, spin-polarized Fermi vapour. These effects are very
weak at very low temperatures and the vapour may be treated as an ideal
Fermi gas \cite{butts,wallis,stoof_varenna,maddalena}. In the
two-component vapours studied by DeMarco et al. \cite{demarco}, however, s-wave scattering is operative between pairs of $^{40}$K atoms in
different hyperfine states. They have thus been able to measure the s-wave
scattering length of $^{40}$K, to observe directly the p-wave energy threshold
law and to evaporatively cool the vapour down to 5 mK. While the s-wave
scattering determined in this way for $^{40}$K is repulsive (i.e. is described
by a positive scattering length), a negative scattering length for $^6$Li
atoms holds promise of achieving a superfluid state in a mixture of $^6$Li
atoms prepared in two hyperfine states \cite{stoof_marianne}. The insurgence of
superfluidity may be revealed through the study of the elementary
excitations of the vapour \cite{hulet_bcs,baranov,bruun_bcs}.

	In the present work we extend to two-component interacting Fermi
vapours in the normal (non-superfluid) state our former study of the
small-amplitude excitationsof density fluctuations in an ideal Fermi gas
confined in a harmonic trap at zero temperature \cite{ilaria}. We make use of an
equations-of-motion approach which is formulated in full generality in Sect.~2
in order to stress the analogies between the problem of present interest
and that of spin dynamics and spin diffusion in a homogeneous Fermi liquid
in a given state of partial spin polarization \cite{abrikosov,caccamo}. The nature of the
assumptions which are adopted in our further calculations on confined Fermi
vapours is made more explicit and justified by this discussion.

	In Sect.~3 we assume complete equilibrium for the two-component vapour
under spherical harmonic confinement and evaluate the Thomas-Fermi
ground-state densities upon relating the components of the kinetic stress
tensor to the local densities by the ideal-gas formula. We give specific
attention to three different cases, i.e. (i) the $^{40}$K system studied by
DeMarco et al. \cite{demarco}, (ii) a strong-coupling regime in which the repulsive
interactions between the two components of a symmetric vapour drive their
spatial separation, and (iii) a simplified description of the weak-coupling
regime in a symmetric vapour. By a symmetric vapour as treated in (ii) and
(iii) we mean equal numbers of particles in the two components as well as
equal masses and equal confinements, as is relevant in relation to the
experiments of DeMarco et al. \cite{demarco}. The form obtained in (iii) for the
density profile is used in Sect.~4 to obtain an analytic determination of the
eigenvectors and of the dispersion relation for both in-phase and
out-of-phase oscillations in a symmetric vapour in the collisional regime.
The role of the interactions in comparison with the vibrational properties
of an ideal one-component Fermi gas is of main interest here. Finally, Sect.~5
gives a brief summary of our main results and offers some concluding
remarks.

\section{Generalized quantum hydrodynamics in a two component Fermi
fluid}
\label{2st}

	We review in this section some general properties of the dynamics
of a two-component fluid with given equilibrium densities
$n_{\sigma}(\vett r)$, $\sigma$ being a
component index that we shall write as $\sigma=(\uparrow,\downarrow)$
to stress the analogy with the 
problem of spin dynamics in a partially spin-polarized electron gas
\cite{caccamo}. 
The Hamiltonian describing the fluid in the presence of external scalar
potentials $V_{\sigma}(\vett r,t)$ is
\bea
H&=&\sum_{\sigma}\int d^3 r \, \hat \psi^{\dag}_\sigma (\vett r,t)
\left[- \frac{\hbar^2}{2m_{\sigma}} \nabla_{\vett
r}^2+V_{\sigma}(\vett r,t)\right] \hat \psi^{}_\sigma (\vett
r,t)\nonumber \\ &+& \frac 1 2 \sum_{\sigma,\sigma'}\int d^3 r\int d^3
r' \phi_{\sigma,\sigma'}(\vett r,\vett r')\hat \psi^{\dag}_\sigma
(\vett r,t) \hat \psi^{\dag}_{\sigma'} (\vett r',t) \hat
\psi^{}_{\sigma'} (\vett r',t) \hat \psi^{}_\sigma (\vett r,t)\;,
\label{2.1t}
\eea
where $ \hat \psi^{}_\sigma (\vett r,t)$ are the field operators and
$\phi_{\sigma,\sigma'}(\vett r,\vett r')$  the interatomic potentials.
Redistributions of population in the two states are not allowed.

The equations of motions for the partial particle densities
$n_{\sigma}(\vett r,t)$ are
obtained by a
standard procedure (see e.g. \cite{fislett}), involving (i) the derivation of the
equation of motion for the density matrix $\rho_{\sigma}(\vett x,\vett
x';t)=\langle \hat \psi^{\dag}_\sigma (\vett r,t) \hat \psi^{}_\sigma
(\vett r,t)\rangle$  from the Hamiltonian~\refeq{2.1t}, and
(ii) projection on the diagonal $\vett r=(\vett x+ \vett x')/2$. 
Setting $\vett r'=\vett x-\vett x'$, the result is
\bea
m_\sigma \frac{\partial^2 n_\sigma(\vett r,t) }{\partial t^2}&=&
\nabla_{\alpha}^{(\vett r)}\nabla_{\beta}^{(\vett r)}
\Pi^\sigma_{\alpha\beta}(\vett r,t)+\nabla_{\alpha}^{(\vett
r)}\left[ n_\sigma(\vett r,t)\nabla_{\alpha}^{(\vett
r)}V_\sigma^H(\vett r,t)\right]\nonumber \\&+& \sum_{\sigma'}\int
d^3r'\nabla_{\alpha}^{(\vett r)}\left\{\left[\nabla_{\alpha}^{(\vett
r)}\phi_{\sigma,\sigma'}(\vett r,\vett
r')\right]\langle\rho_\sigma(\vett r,t)\rho_{\sigma'}(\vett r',t)
\rangle_c\right\}\;,
\label{2.2t}
\eea
where the convention of summation over repeated Cartesian indices in the
derivatives has been adopted. In Eq.~\refeq{2.2t} we have defined the kinetic
stress tensors
\be
\Pi^\sigma_{\alpha\beta}(\vett
r,t)=-\frac{\hbar^2}{m_{\sigma}}\nabla_{\alpha}^{(\vett 
r')}\nabla_{\beta}^{(\vett
r')}\left.\rho_{\sigma}(\vett{r-r'}/2,\vett{r+r'}/2;t) \right|_{\vett r'=0}
\label{2.3t}
\ee
and the mean-field potentials
\be
V_\sigma^H(\vett r,t)=V_\sigma(\vett r,t)+\sum_{\sigma'}\int d^3r'\,
\phi_{\sigma,\sigma'}(\vett r,\vett r') n_{\sigma'}(\vett r',t)  \;.
\label{2.4t}
\ee
The non-mean-field effects are collected in the last term on the RHS
of Eq.~\refeq{2.2t}, where $\rho_{\sigma}(\vett r,t)$ is the density operator and $\langle\rho_\sigma(\vett r,t)\rho_{\sigma'}(\vett r',t)
\rangle_c$ is the cluster part of the
density-density correlations. No assumption has as yet been made on the
temperature of the fluid.

The equilibrium equations determining the density profiles
$n_\sigma(\vett r)$ are
obtained from Eq.~\refeq{2.2t} by taking the static limit. The
equations of motion 
for the density fluctuations $\delta n_\sigma(\vett r,t)$ driven by
weak external potentials are then 
obtained by writing $n_\sigma(\vett r,t)=n_\sigma(\vett r)+\delta n_\sigma(\vett r,t)$ in Eq.~\refeq{2.2t} and by linearizing it. We shall go
through these steps in  Section~\ref{3st} and~\ref{4st} for a dilute
two-component Fermi vapour. 
Here we proceed to introduce the approximations that we shall make in the
dynamical treatment of Sect.~\ref{4st} by discussing the
non-mean-field term in Eq.~\refeq{2.2t}. 

\subsection{Interdiffusion in the two component fluid}
\label{2.1st}

	We evaluate in this section the role of collisions between
fluctuations in determining damping of collective motions in the
two-component fluid. As a preliminary we recall that the linearized
equations of motion for the partial density fluctuations in the
two-component fluid are conveniently transformed into those for the total
particle density fluctuation $\delta n (\vett r,t)$ and for the concentration fluctuation $\delta M (\vett r,t)$ (the
"magnetization" fluctuation in the electron gas analogue)
by taking simple linear combinations of the two $\delta n_\sigma(\vett
r,t)$'s (see e.g.\cite{kim}). 

	We consider first the dynamics of small fluctuations around a
homogeneous equilibrium state, where we can appeal to the treatment given
by Caccamo et al. \cite{caccamo} for the evaluation of the interdiffusion (or "spin
diffusion") coefficient. Momentum conservation ensures that the only
non-vanishing inverse relaxation time in the hydrodynamic limit is the
interdiffusion one, say $\tau_{MM}^{-1}$, which is written as
\be
\tau_{MM}^{-1}=n \gamma_M/(m n_{\uparrow}n_{\downarrow})
\label{2.5t}
\ee		
where $n_{\uparrow}$ and $n_{\downarrow}$ are the partial equilibrium
densities, $n=n_{\uparrow}+n_{\downarrow} $ is the total density
and we have assumed $m=m_{\uparrow}=m_{\downarrow}$. 
An exact expression for the quantity $\gamma_M$ in Eq.~\refeq{2.5t}
is obtained from the non-mean-field term in Eq.~\refeq{2.2t} in the appropriate
hydrodynamic limit. In the case of a central pair potential this reads
\be
\gamma_M=\int d^3r \, (\hat k \cdot
\nabla)\phi_{\uparrow\downarrow}(r)\frac{\partial\langle\rho_\uparrow(\vett
R)\rho_{\downarrow}(\vett R +\vett r) 
\rangle_c}{\partial v_{\uparrow}}
\label{2.6t}
\ee
in a reference frame where the component $\uparrow$ is at rest and the other
component is flowing with a uniform drift velocity $v_{\downarrow}$.

	The Fourier transform of the non-equilibrium correlation function
in Eq.~\refeq{2.6t} is evaluated in a binary collision approximation by the
decoupling procedure used by Baym \cite{baym_old} in treating the electrical
resistance of metals (see also Kadanoff and Baym \cite{libro_kadanoff}). Namely,
\bea
\lefteqn{\Im {\rm F.T.} \left\{\langle \rho_\uparrow(\vett
R)\rho_{\downarrow}(\vett R +\vett r) 
\rangle_c\right\}_{\vett k}\nonumber =}\\
&&\frac 1 2 n \hbar
\phi_{\uparrow\downarrow}(k)\int_{-\infty}^{\infty}\frac{d\omega}{2
\pi}\left[\tilde S_{\uparrow \uparrow}(\vett k, \omega)\tilde S_{\downarrow \downarrow}(-\vett k, -\omega) -\tilde S_{\uparrow \uparrow}(-\vett k, -\omega)\tilde S_{\downarrow \downarrow}(\vett k, \omega)\right]\;,
\label{2.7t}
\eea
where $\tilde S_{\sigma\sigma}(\vett k \omega)$ is the van Hove dynamic
structure factor of each component in the 
non-equilibrium state. For the dilute Fermi fluid of present interest we
can replace the interaction potential in Eq.~\refeq{2.7t} by a contact interaction
and  $\tilde S_{\sigma\sigma}(\vett k \omega)$  by the ideal-gas value corresponding to a displaced Fermi sphere for
the $\downarrow$ component. Following the lines of the calculation
given in Ref.~\cite{caccamo} 
and taking for simplicity $n_{\uparrow}=n_\downarrow$ in
Eq.~\refeq{2.5t}, we find to leading order in the 
temperature $T$ the result
\be
\tau_{MM}^{-1}=(4 \pi m a_{\uparrow\downarrow}^2E_F^2/3\hbar^3)(T/T_F)^2\;.
\label{2.8t}
\ee
Here, $a_{\uparrow\downarrow}$ is the (triplet) scattering length, $E_F$ is the Fermi energy and $T_F=E_F/k_B$.
This result
could also be obtained directly from Eq. (6.8) in Ref.~\cite{caccamo} upon replacing
a screened Coulomb
interaction by a contact interaction.

	The result given in Eq.~\refeq{2.8t} above for a homogeneous,
two-component Fermi fluid can now be used for an estimate of the role of
collisions in a confined Fermi fluid. We replace the Fermi energy $E_F$ by its
local value, which in the case of harmonic confinement in a spherical trap
characterized by a frequency $\omega_f$ is
\be
E_F=(3N)^{1/3}\hbar \omega_f
\label{2.9t}
\ee
with $N$ the total number of fermions. Hence,
\be
(\omega_f\tau_{MM})^{-1}=(4\pi/3^{1/3}) (N^{1/3}
a_{\uparrow\downarrow}/a_{ho})^2(T/T_F)^2  
\label{2.10}
\ee
$a_{ho}=(\hbar/m\omega_f)^{1/2}$ being the harmonic-oscillator
length. A similar result has been reported 
recently by Vichi and Stringari \cite{lorenzo} from a collision-integral
approach.

	In summary, because of momentum conservation the damping processes
in the hydrodynamic limit of a two-component Fermi fluid are associated
with collisions between the two components and affect only their relative
motions. These processes vanish quadratically with decreasing temperature
because of Fermi statistics (see also \cite{abrikosov}). A collisional regime may
nevertheless be established by scattering against impurities (see for
instance the work of Ruckenstein and L\'evy \cite{ruck} on spin dynamics in
paramagnetic quantum fluids). We turn below to an estimate of these
collision processes in the normal Fermi fluid of present interest.

\subsection{Collisional regime via impurity scattering}
\label{2.2st}

A collisional regime is established in the low-temperature vapour
{\it for both in-phase and out-of-phase modes of motion of the two components}
when the inequality
\be
\omega\tau \ll 1
\label{2.11t}
\ee
holds, $\tau$ being the collision time for scattering of fermions against
impurities and $\omega$ being on the scale of the trap frequency
($\omega \simeq \omega_f$) for
low-frequency modes.

	For an estimate of the needed number $N_s$ of scatterers we take the
impurities as cold atoms with a mean velocity which is negligible relative
to that of the fermions. We can then write $\tau=l/v$, $l$ and $v$
being the mean free 
path and the average speed of a fermion. We have $l=\Sigma^{-1}$,
where $\Sigma$ is the 
macroscopic cross-section given by $\Sigma=n_s \sigma$ in terms of the
density $n_s $ of
scatterers and of the cross-section $\sigma$ for fermion-impurity scattering
(see for instance Ref.~\cite{weinberg}). Setting $n_s=N_s(4\pi
a_{ho}^3/3)^{-1}$ 
and $\sigma=4 \pi a_{sc}^2$ with $a_{sc}$ the fermion-impurity
scattering length, and taking $v=(3E_F/4m)^{1/2}$ with $E_F$ given by
Eq.~\refeq{2.9t}, we find 
\be
\omega_f\tau=\frac{(4/27)^{1/2}}{N_s (3N)^{1/6}}\left(\frac{a_{ho}}{a_s}\right)^2
\label{2.12t}
\ee
	
For illustrative purposes we consider the case of $^{39}$K or $^{41}$K
bosonic impurities in the gas of $^{40}$K fermions studied by DeMarco
et al. \cite{demarco} 
($N\simeq 10^7$ and $\omega_f\simeq 209$ s$^{-1}$, the latter being the geometric mean of the radial and axial
frequencies in the experiment). From the known values of the $^{39}$K-$^{40}$K and
$^{41}$K-$^{40}$K scattering lengths ($a_s \simeq$3600 and $a_s\simeq$  93 Bohr radii, respectively)
we find that a number of $^{39}$K impurities of order $N_s\simeq 10^{-6}N$, or of $^{41}$K impurities
of order $N_s \simeq 10^{-3}N$, would suffice to verify the
inequality~\refeq{2.11t}  with $\omega=\omega_f$.

	We conclude, therefore, that a collisional regime can easily be
established for the low-frequency excitations of trapped Fermi vapours.
This regime reflects rather directly the quantal statistics of the vapour
\cite{ilaria} and we study it for the two-component Fermi fluid in
Sect.~\ref{4st}
 below.
Excitations in the collisionless regime are of less interest, since they
mostly reflect the frequency of the trap \cite{ilaria,lorenzo}.

\section{Equilibrium density profiles in spherical confinement}
\label{3st}

As already discussed in Sect.~\ref{1st}, we treat a dilute two-component Fermi
gas at zero temperature in which only $s$-wave scattering between pairs of
fermions in different hyperfine states is operative. This coupling is
described by the parameter $f=4 \pi \hbar^2
a_{\uparrow\downarrow}/m$. In the experimentally relevant situation the 
two populations have not only the same mass but also essentially identical
numbers and trap frequencies. However, we shall impose the equality
$N_{\uparrow}=N_{\downarrow}=N/2$ only 
later below.

	We take the gas as being statically in the equilibrium state and
dynamically in the collisional regime. As in our earlier work \cite{ilaria}, we
relate the kinetic stress tensor of each component to its local density by
the homogeneous Fermi gas formula,
\be
\Pi_{\alpha\beta}^\sigma(\vett r,t)= \delta_{\alpha\beta} \frac{2}{5}
A [n_\sigma(\vett r,t)]^{5/3}
\label{3.1t}
\ee
where $A= \hbar^2 (6 \pi^2)^{2/3}/2m$. Such a local density approximation assumes that the length scale
for the variation of the density profiles in space is large relative to the
inverse Fermi wave number $k_{f}^{-1}$ and to the length $c/\omega$, with $\omega$ the excitation
frequency and $c$ the velocity of sound propagation in the homogeneous
fluid.

	The equilibrium density profiles are then easily obtained from the
static limit of Eq.~\refeq{2.2t} in the mean-field approximation. They have the
Thomas-Fermi form,
\be
n_\sigma(\vett r)=\theta [\epsilon_\sigma-V_\sigma(\vett r)- f n_{\bar
\sigma}(\vett r)] \left\{ A^{-1}[\epsilon_\sigma-V_\sigma(\vett r)- f n_{\bar
\sigma}(\vett r)]\right\}^{3/2}\;,
\label{3.2t}
\ee
where $\bar \sigma$ denotes the component different from $\sigma$. 
In Eq.~\refeq{3.2t} $V_\sigma(\vett r)$ are the
static confining potentials and $\epsilon_\sigma$ are the chemical
potentials, to be 
determined from the condition $N_{\sigma}=\int d^3r \,n_\sigma(\vett r)$. We emphasize that the $N_{\sigma}$'s are fixed,
{\it i.e.} these equations do not allow for redistributions of
population in the 
two hyperfine states.

	In the same approximation the total energy of the vapour is the sum
of three terms, {\it i.e.} a kinetic energy $E_{kin}$, a potential
energy $E_{ho}$  and an
interaction energy $E_{int}$. These are
\be
E_{kin}=(6\pi^2)^{2/3}\frac{3 \hbar^2}{5 m}\sum_{\sigma}\int d^3r
\left[n_\sigma(\vett r)\right]^{5/3}
\;,
\label{3.3t}
\ee
\be
E_{ho}=\sum_{\sigma}\int d^3r
\,n_\sigma(\vett r)V_\sigma(\vett r)
\label{3.4t}
\ee
and
\be
E_{int}=\sum_{\sigma}\int d^3r
\,n_\uparrow(\vett r)n_\downarrow(\vett r)
\label{3.5t}
\ee
These will be helpful in understanding the behaviour of the vapour at
strong coupling in Sect.~\ref{3.2st}.

\subsection{An illustrative example for a weakly coupled $^{40}$K vapour}

\begin{figure}
\centerline{\psfig{figure=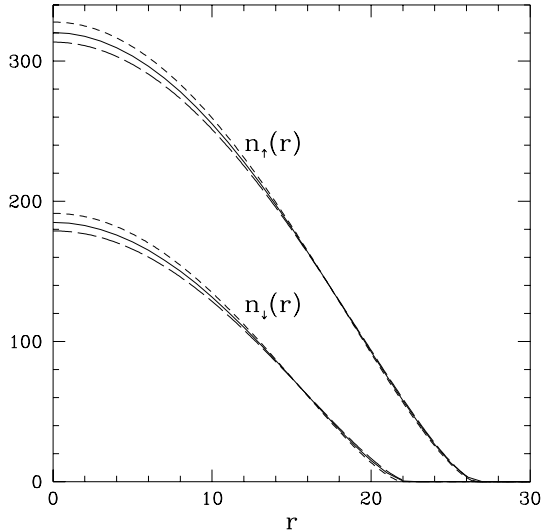,width=0.5\linewidth}}
\caption{Density profiles $n_\uparrow(r)$ and  $n_\downarrow(r)$ (in
units of $a_{ho}^{-3}$) versus distance $r$ from
the centre of a spherical trap (in units of $a_{ho}$) in a mixture of
$10^7$ fermions 
at composition $N_\uparrow/N_\downarrow=3$. The long (short) dashed
curves show the profiles 
corresponding to an $s$-wave scattering length $a_{\uparrow\downarrow}=157\,a_B$ ($a_{\uparrow\downarrow}=-157\,a_B$), relatively to the case
of the ideal mixture (full curves).}
\label{1f}
\end{figure}

	Figure~\ref{1f} reports the numerical results that we obtain from Eq.~\refeq{3.2t} for the density profiles in a gas subject to spherical harmonic
confinement, with system parameters chosen after the experiment of DeMarco
{\it et al.} \cite{demarco}
($\omega_f=(\omega_{\parallel}\omega_{\perp}^2)^{1/3}=209$ s$^{-1}$,
$N=10^7$ and $a_{\uparrow\downarrow}=157$ Bohr radii) but at
composition $N_{\uparrow}/N_{\downarrow}=3$. The cases
$a_{\uparrow\downarrow}=-157\,a_B$ and $a_{\uparrow\downarrow}=0$
 are also shown.

	Evidently, the effects of the interactions are small in this
situation and very simple to understand: a repulsion (attraction) between
the two components disfavours (favours) their
overlap in the central part of the trap.

\subsection{ Spatial separation of the two components at strong
coupling}
\label{3.2st}

\begin{figure}
\centerline{\psfig{figure=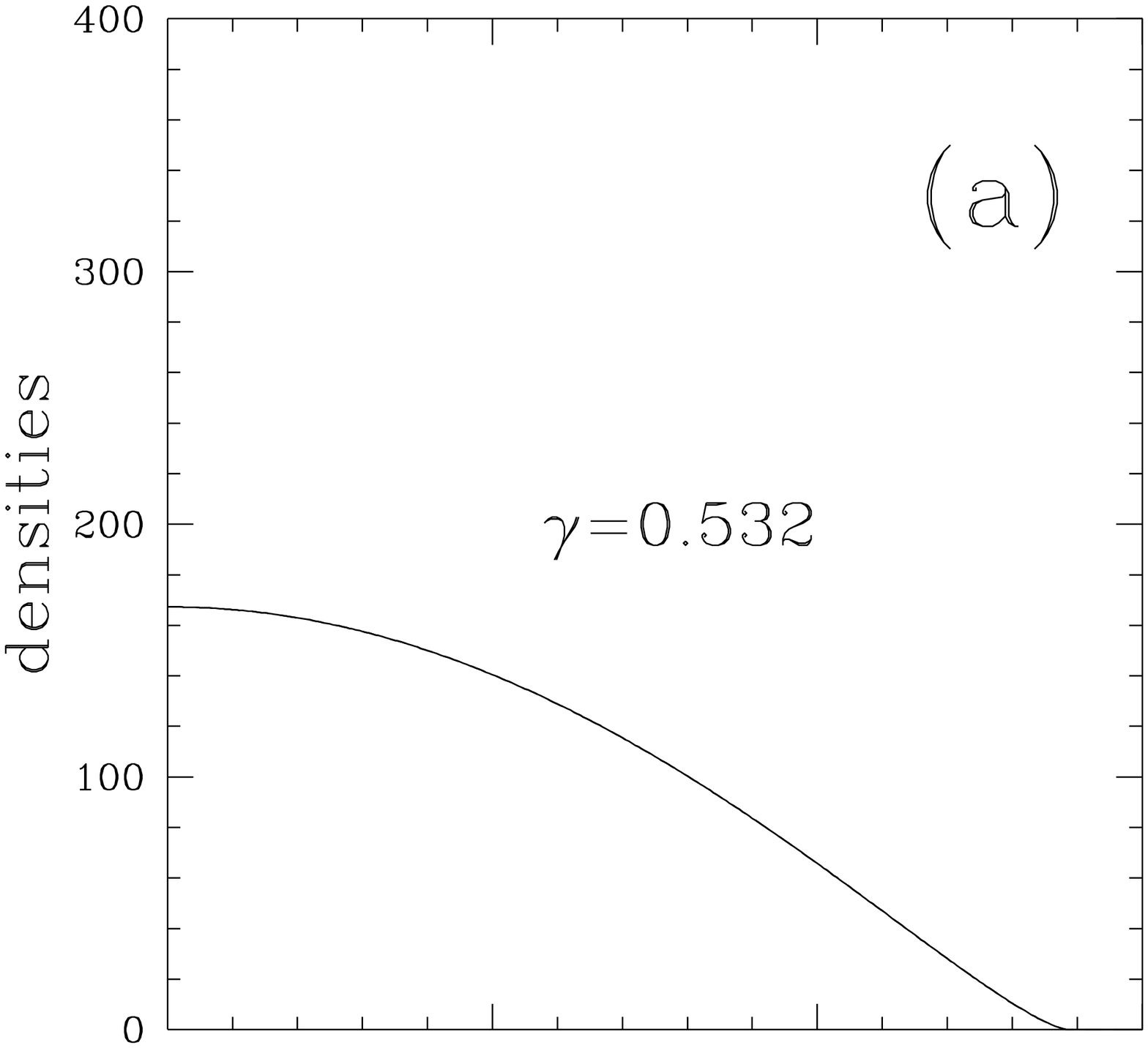,width=0.5\linewidth}\psfig{figure=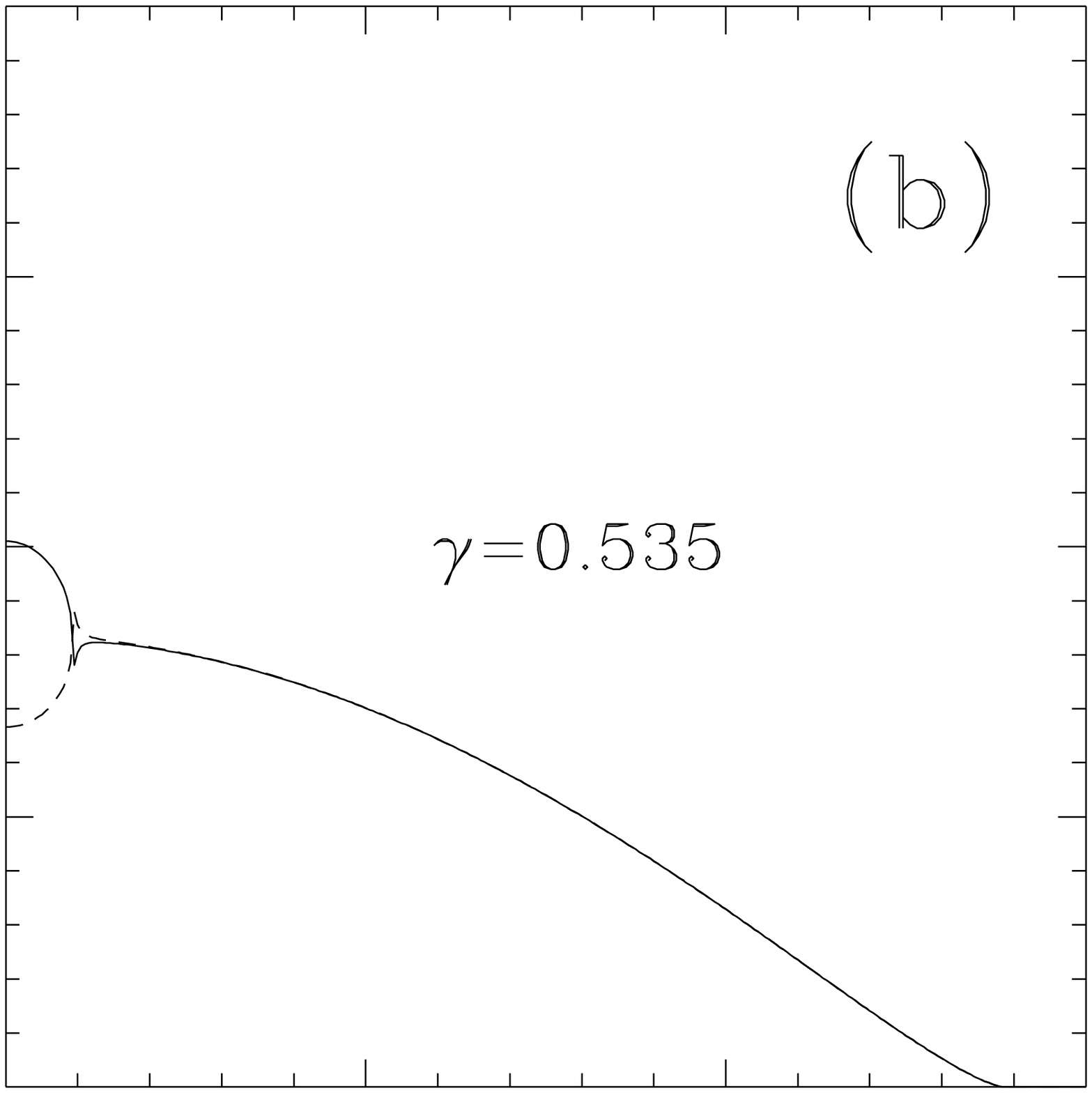,width=0.5\linewidth}}
\centerline{\psfig{figure=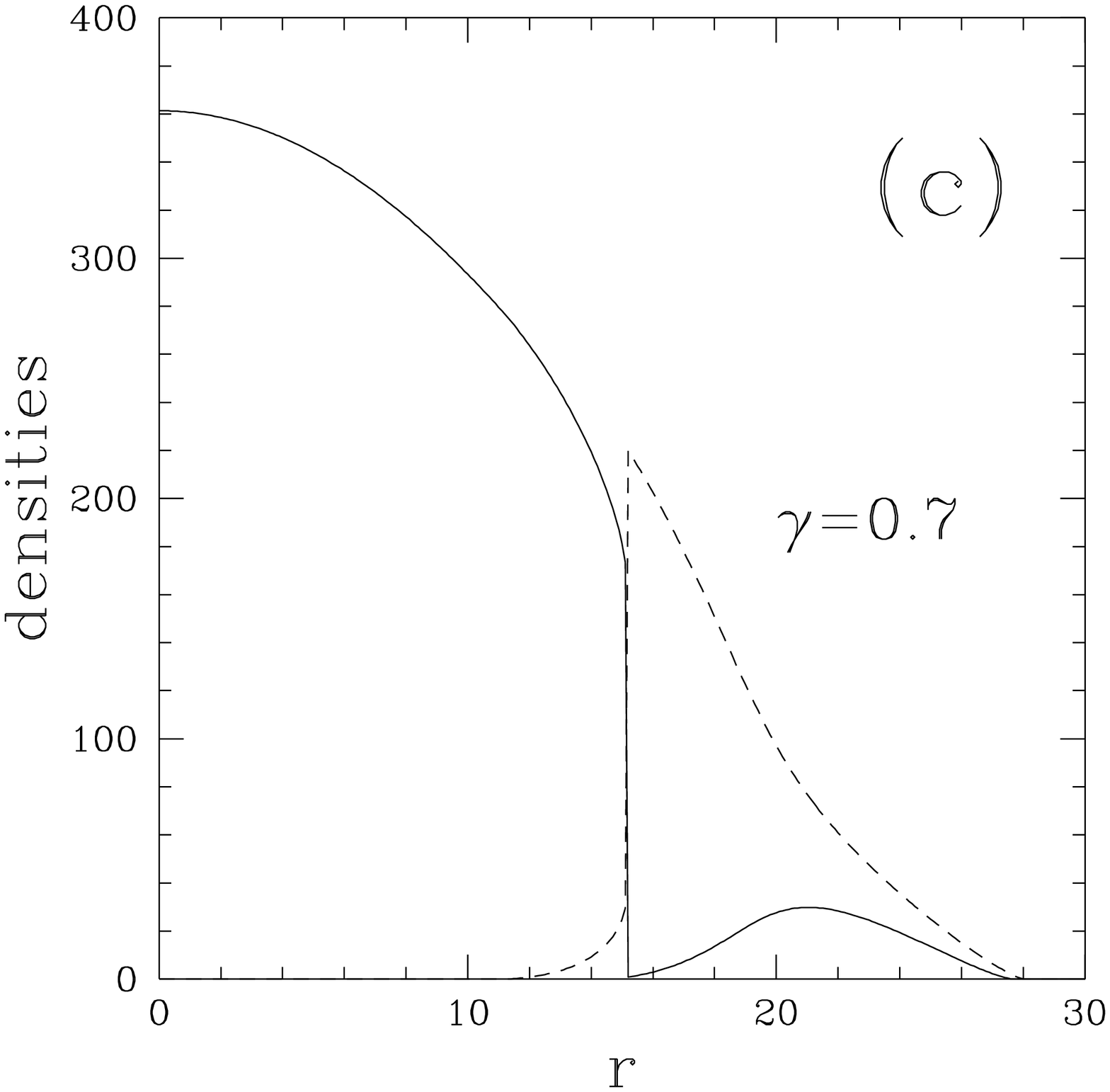,width=0.5\linewidth}\psfig{figure=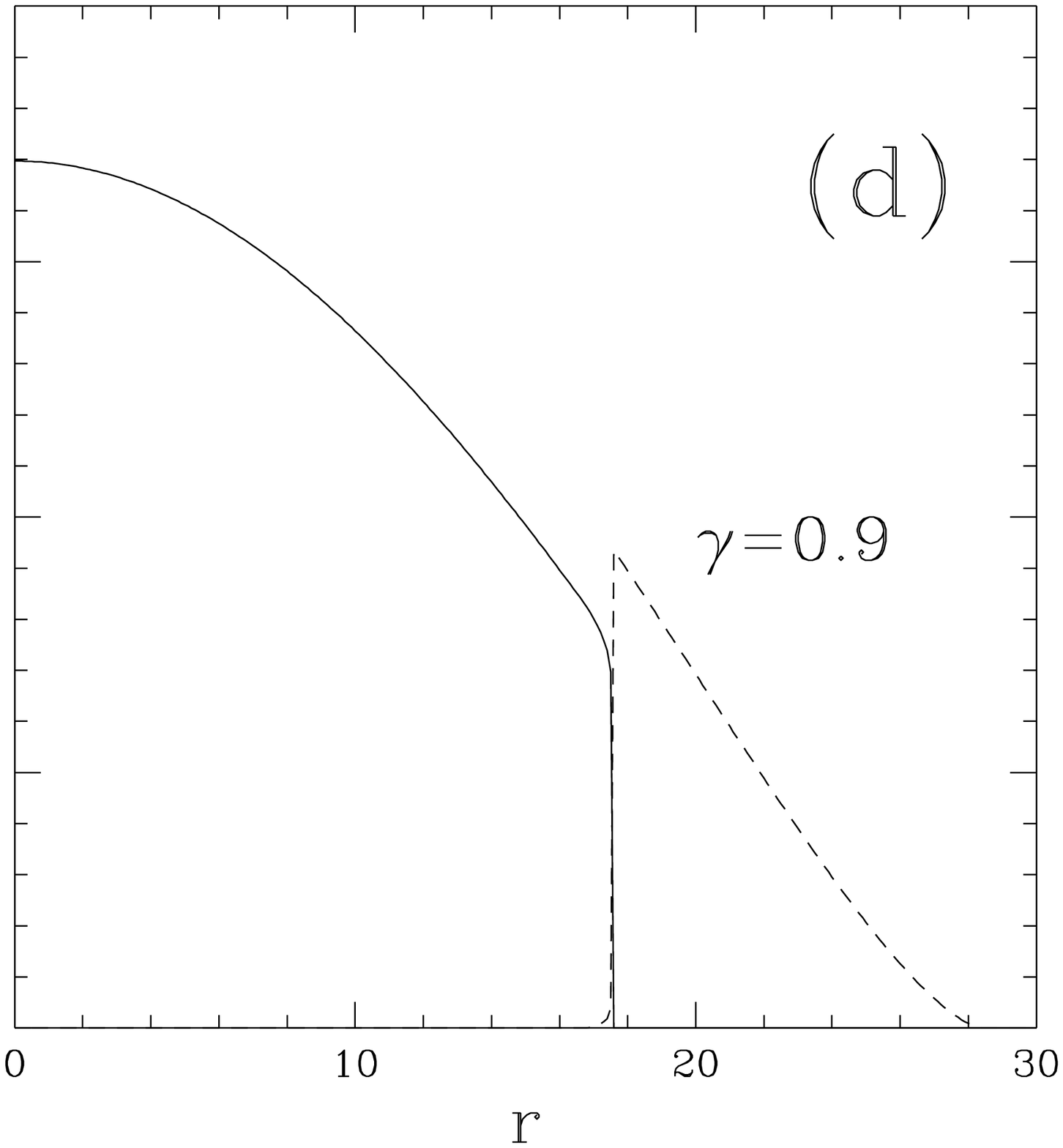,width=0.5\linewidth}}
\caption{Density profiles (in units of $a_{ho}^{-3}$) versus distance
$r$  from the
centre of a spherical trap (in units of $a_{ho}$) in a symmetric mixture of
fermions, at various values of the coupling strength parameter
$\gamma$. In (a), 
the profiles of the two components are still in complete overlap. Spatial
symmetry breaking is first visible in (b). Separation of the two components
(shown by full and dashed curves) continues through (c) and (d).}
\label{2f}
\end{figure}

\begin{figure}
\centerline{\psfig{figure=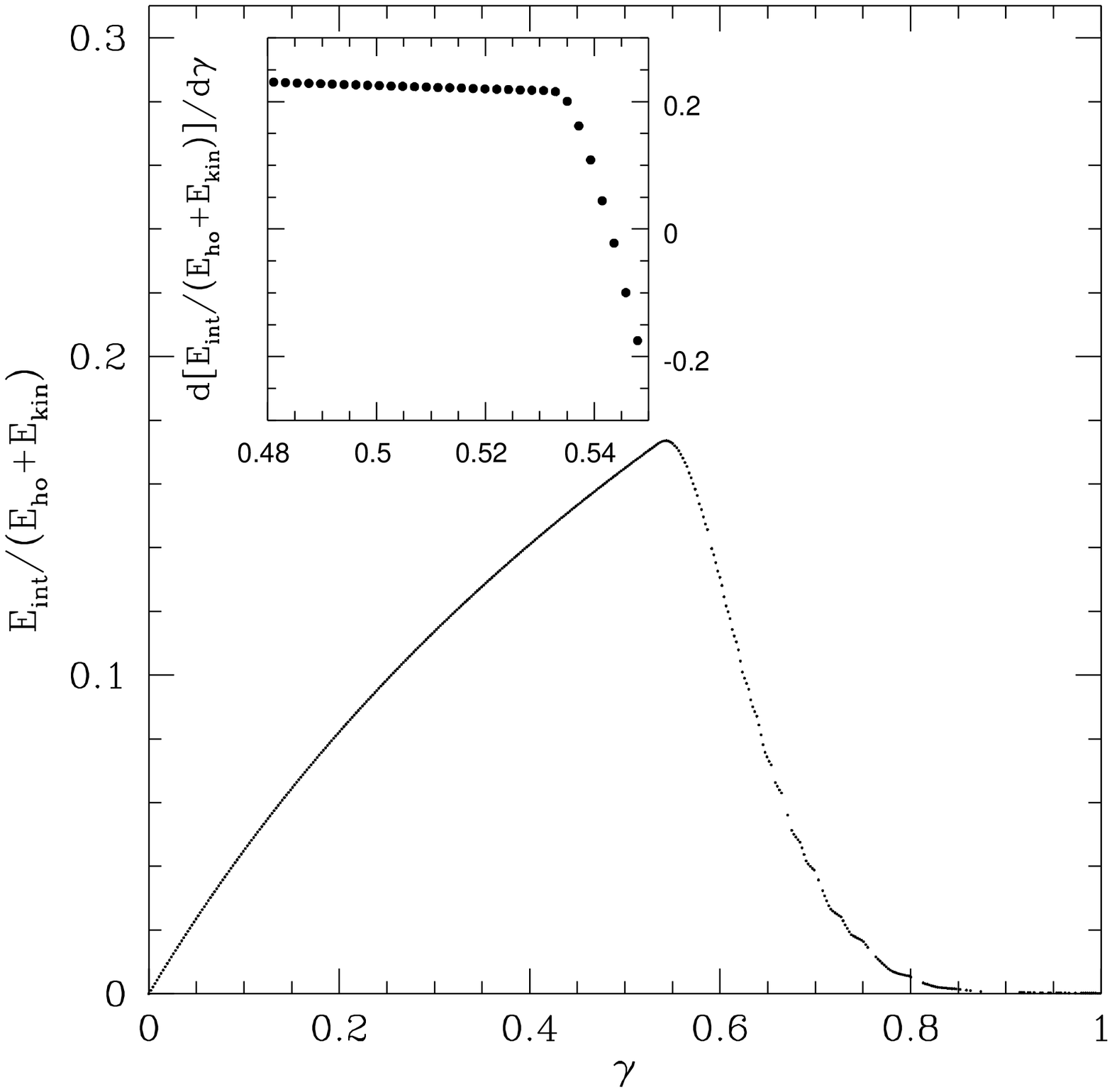,width=0.5\linewidth}}
\caption{The ratio of the interaction energy $E_{int}$ to the sum
$E_{ho}+E_{kin}$ of the 
harmonic-oscillator and kinetic energies, plotted against the coupling
strength $\gamma \simeq 0.5 N^{1/6}(a_{\uparrow\downarrow}/a_{ho})$
for a symmetric mixture of fermions in spherical confinement. 
The inset shows the first derivative of the same function.}
\label{3f}
\end{figure}

	We may expect that in the case of repulsive interactions, with
increasing coupling strength and still barring transitions between the two
hyperfine levels as already noted under Eq.~\refeq{3.2t}, the gas will
be led to 
diminish its total energy by reducing the spatial overlap between the two
components. Figure~\ref{2f} shows how such symmetry breaking occurs in
a spherical 
trap, for $N_\uparrow=N_\downarrow$
 and under the condition that overall spherical symmetry be
maintained. In this case one component is pushed away from the centre of
the trap and the gas configuration becomes that of a central core enriched
in one component and surrounded by a spherical shell enriched in the
other.

	The symmetry breaking is driven by the competition between the
repulsive interaction energy, favouring spatial separation of the
components, and the kinetic energy disfavouring localization. Evidently and
in contrast to the behaviour illustrated in Figure~\ref{2f}, a gas confined in an
axially symmetric trap will tend to reduce its energy {\it via} relative shifts
in the centres of the two clouds. A rich phase diagram will ensue if "spin
flips" between the two hyperfine states are also allowed. In the following
we estimate the critical coupling strength at which spatial symmetry
breaking occurs in terms of the number of fermions (or alternatively of the
$a_{\uparrow\downarrow}$ scattering length) for the case illustrated
in Figure~\ref{2f}.

	The four cases of density profiles illustrated in
Figure~\ref{2f} are
labelled by a parameter $\gamma$, which is defined by
\be
\gamma =\alpha N^{1/6}(a_{\uparrow\downarrow}/a_{ho})
\label{3.6t}
\ee	
with $\alpha=2^{1/3}3^{1/6}(8192/2835\pi^2)$ \cite{lorenzo}. In fact the value
of $\gamma$ in Eq.~\refeq{3.6t} is obtained as
$\gamma=[E_{int}/(E_{ho}+E_{kin})]_0$ when the 
ratio $E_{int}/(E_{ho}+E_{kin})$ is calculated from the density
profile of the Fermi gas in the 
absence of interactions. Figure~\ref{3f} reports the true values of
$E_{int}/(E_{ho}+E_{kin})$ against $\gamma$
for the system described in Figure~\ref{2f}. The (obvious) linear shape of this
function at weak couplings gently bends over with increasing coupling,
until an almost sharp break occurs at spatial symmetry breaking. This is
emphasized in the inset in Figure~\ref{3f} giving the derivative of $E_{int}/(E_{ho}+E_{kin})$ with respect
to $\gamma$. We have checked that the same plot is obtained for $E_{int}/(E_{ho}+E_{kin})$ by varying $a_{\uparrow\downarrow}/a_{ho}$ at
constant $N$ and by varying $N$ at constant
$a_{\uparrow\downarrow}/a_{ho}$.

	In the experiments of DeMarco et al. \cite{demarco} on $^{40}$K with
$N\simeq 10^7$, the value
of $\gamma$ is $\gamma\simeq 0.022$ i.e. still very far from the
critical value $\gamma_c\simeq 0.535$ for the symmetry
breaking illustrated in Figures~\ref{2f} and~\ref{3f}.
 The weak dependence of $\gamma$ on $N$ in
Eq.~\refeq{3.6t} implies that a number of $^{40}$K atoms of order
$10^{15}$ would have to be 
reached if all other system parameters remain the same. A parallel increase
in the ratio $a_{\uparrow\downarrow}/a_{ho}$ as suggested by
Eq.~\refeq{3.6t} would evidently be helpful in 
relaxing such stringent condition.

\subsection{Approximate form of the density profile for a symmetric
vapour at weak coupling}
\label{3.3st}

\begin{figure}
\centerline{\psfig{figure=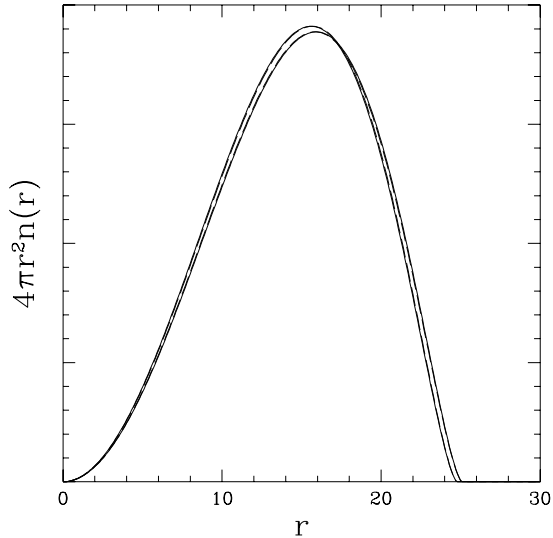,width=0.5\linewidth}}
\caption{Illustrating the accuracy of the approximate form~\refeq{3.8t} of the
particle distribution $4 \pi r^2 n(r)$ (dashed curves) relative to the
full Thomas-Fermi 
profile (full curves), for a symmetric mixture of fermions in the two cases
$\gamma=\pm 0.022$.}
\label{4f}
\end{figure}

In the case $N_\uparrow=N_\downarrow$ the shape of the total
density profile $n(\vett r)=n_\uparrow(\vett r)+n_\downarrow(\vett r)$ at weak 
coupling is well represented by a form which is suitable for the analytic
study of the eigenmodes of the gas that we report in Sect.~\ref{4st}
below. From Eq.~\refeq{3.2t} the Thomas-Fermi density profile is
\be
n(\vett r)= 2 A^{-3/2}\left[\epsilon_F-\frac 1 2 m \omega_f^2 r^2
-\frac f 2 n(\vett r)\right]^{3/2}
\label{3.7t}
`\ee
for $r\leq R_F$, where $R_F=(2 \epsilon_F/m \omega_f^2)^{1/2}$ and $\epsilon_F=\epsilon_\uparrow=\epsilon_\downarrow$ is the chemical potential of the mixture. At weak
coupling the profile~\refeq{3.7t} can be approximated by
\be
n(\vett r)=\frac{8 N}{\pi^2 R_F^3}(1-r^2/R_F^2)^{3/2}\theta(1-r^2/R_F^2)\;,
\label{3.8t}
\ee
where $\epsilon_F$ and $R_F$ still are the true chemical potential and
the Fermi radius in 
the interacting mixture. The form~\refeq{3.8t} is adjusted to preserve
normalization to $N$ as well as the value of $R_F$.

	Figure~\ref{4f} compares the approximate form~\refeq{3.8t}
with the correct 
Thomas-Fermi form~\refeq{3.7t} at $\gamma=\pm 0.022$ by plotting the
function $4 \pi r^2 n(r)$. The phase-space
factor $4 \pi r^2$ masks the small differences that would be present
in the two forms 
of $n(\vett r)$ near the centre of the trap. On the other hand,
preserving the correct 
value of the Fermi radius in the approximate form~\refeq{3.8t} is
crucial in view 
of the boundary conditions to be imposed in the determination of the
eigenmodes of the vapour.

	As a final remark we notice that the profile in
Eq.~\refeq{3.8t}  has the
same form as for an ideal one-component Fermi gas \cite{molmer}. This fact is
crucial for the analytic treatment of the dynamics of density fluctuations
in a weakly coupled symmetric mixture, that we give in the next
section.

\section{Dynamics of density fluctuations}
\label{4st}

The equations of motions~\refeq{2.2t}, after linearization in
the partial 
density fluctuations and adopting (i) the mean-field approximation
($\langle\rho_{\sigma}(\vett r,t)\rho_{\sigma'}(\vett r',t)\rangle_c = 0$)
and (ii) the local density approximation for the kinetic stress tensor (Eq.~\refeq{3.1t}, reduce to
\bea
m \frac{\partial^2 n_{\sigma}(\vett r)}{\partial t^2}&=&\nabla^2 \left[\frac 2
3 A n_\sigma^{2/3}(\vett r) \delta n_\sigma (\vett r,
t)\right]\nonumber \\ &+& \nabla \cdot \left\{\delta n_\sigma (\vett r,
t) \nabla[V_{\sigma}(\vett r)+ f n_{\bar \sigma} (\vett r,
t) ]+ f n_\sigma (\vett r)\nabla\delta
n_{\bar \sigma} (\vett r,
t) \right \}
\label{4.1t}
\eea
at resonance ({\it i.e.} for $V_{\sigma}(\vett r,t)=V_{\sigma}(\vett
r)$). With the help of the equilibrium conditions~\refeq{3.2t}, and taking Fourier transforms with respect to the time variable, Eq.~\refeq{4.1t}
can be written as
\bea
-m \omega^2 \delta n_\sigma (\vett r, \omega)=\frac 1 3 A \left[2
 n_{\sigma}^{2/3} (\vett r) \nabla^2 +\nabla(n_{\sigma}^{2/3})\cdot
 \nabla -\nabla^2(n_{\sigma}^{2/3})\right]\delta n_\sigma (\vett r,
 \omega)\nonumber \\ + \left\{ [\epsilon_{\bar \sigma}-V_{\bar
 \sigma}(\vett r)- A n_{\bar \sigma}^{2/3} ] \nabla^2 - \nabla[V_{\bar
 \sigma}(\vett r)+ A n_{\bar \sigma}^{2/3} ] \cdot \nabla
 \right\}\delta n_{\bar \sigma} (\vett r,
 \omega)
\label{4.2t}
\eea
Evidently, Eq.~\refeq{4.2t} describes a two-by-two eigenvalue problem for the
coupled partial density fluctuations, which is to be solved numerically in
the general case.

	The problem is considerably simplified in the case of a symmetric
mixture ($m_\uparrow=m_\downarrow$, $N_\uparrow=N_\downarrow$ and
$V_\uparrow(\vett r)=V_\downarrow(\vett r)$, these conditions being
well satisfied in the 
experiments of DeMarco et al. \cite{demarco}). In this case the dynamical
equations~\refeq{4.2t}  lead to separate equations of motion for the
total density 
fluctuations $\delta n(\vett r,\omega)=\delta n_{\uparrow}(\vett r,\omega)+\delta n_{\downarrow}(\vett r,\omega)$ and for the concentration fluctuations $\delta M(\vett r,\omega)=\delta n_{\uparrow}(\vett r,\omega)-\delta n_{\downarrow}(\vett r,\omega)$. The eigenvalue
equation for $\delta n(\vett r,\omega)$ reads
\bea
-m \omega^2 \delta n({\vett r},\omega)=
\nabla \cdot \left(\frac{2}{3}A (n/2)^{2/3} \nabla \delta n(\vett
r,\omega)- \frac{A}{3}  
(\nabla (n/2)^{2/3}) \delta n\right. \nonumber \\ +\left.  (\epsilon_F-V({\vett r})-A (n/2)^{2/3}) \nabla \delta n(\vett r,\omega) 
\right)\;.
\label{4.3t}
\eea
Similarly, the eigenvalue equation for $\delta M(\vett r,\omega)$ is
\bea
-m \omega^2 \delta M({\vett r},\omega)=
\nabla \cdot \left(\frac{2}{3}A (n/2)^{2/3} \nabla \delta M - \frac{A}{3} 
(\nabla (n/2)^{2/3}) \delta M({\vett r},\omega)\right. \nonumber \\
-\left.  (\epsilon_F-V({\vett r})-A (n/2)^{2/3}) \nabla \delta M({\vett
r},\omega) 
\right)\;.
\label{4.4t}
\eea
Equations~\refeq{4.3t} and~\refeq{4.4t} can be solved analytically by
the technique used 
in our earlier work on the ideal one-component Fermi gas \cite{ilaria}, if the
form~\refeq{3.8t} is adopted for the equilibrium profile. As already
discussed in Sect.~\ref{3.3st}, Eq.~\refeq{3.8t} becomes accurate at
small coupling. We shall again impose 
that the solutions vanish continuously at the cloud boundary, as a
consequence of Fermi statistics giving a high cost in kinetic energy to
rapid variations of the densities in space.
	For both in-phase and out-of-phase motions of the two components,
the frequency eigenvalues depend on a parameter $C$ given by
\be
C=(3N)^{2/3}(\hbar \omega_f/\epsilon_F)^2
\label{4.5st}
\ee
This quantity is the square of the ratio of the ideal Fermi energy to the
true Fermi energy and hence, in the case of repulsive interactions where $\epsilon_F$
increases with the scattering length, is limited from above by the
inequality $C<1$. All the mathematical details of the solution of
Eqs.~\refeq{4.3t} and~\refeq{4.4t} are given in
Appendix~\ref{appdf}. Here we report only the main results. 

\subsection{Small oscillations  of total density fluctuations}
	The eigenfunctions of the total density oscillations vanish at the
Fermi radius $r=R_F$ provided $C<3$, in a way which depends on the
parameter $C$ and 
hence on the strength of the interactions (see Appendix~\ref{appdf1} for their
detailed expressions).

	The corresponding eigenfrequencies are labelled by the angular
momentum number $l$ and by an integer $n$ representing the number of internal
nodes in the density fluctuation profile. The dispersion relation is
\be
(\omega_{nl}/\omega_f)^2=l+2n + \frac n 3 (3 -C)(2n +2l +1)
\label{4.6t}
\ee
The ideal Fermi gas limit corresponds to $C=1$ and in this case
Eq.~\refeq{4.6t} 
yields back our earlier result \cite{ilaria}. More generally, the dispersion
relation~\refeq{4.6t}  reduces to that of the ideal Fermi gas only for
the surface 
modes (i.e. for $n=0$). Instead, the frequencies of the modes with $n>0$ are
shifted by the interactions.

	The above results are easily extended to evaluate the low-frequency
modes in an axially symmetric confinement (see e.g. \cite{ilaria}). As
discussed in Sect.~\ref{2st}, damping of these modes will set in when
$\omega \tau \simeq 1$ for scattering against
cold-atom impurities.

\subsection{Small oscillations of concentration fluctuations}
	The eigenfunctions vanish at the Fermi radius only if $C>3/5$ (see
Appendix~\ref{appdf2} for
the details). This inequality marks the breakdown of the present
approximation in the case of repulsive interactions. Under this restriction
the dispersion relation for concentration fluctuations having $n$ internal
nodes is
\be
(\omega_{nl}/\omega_f)^2=(l+2n)(2C-1) + \frac n 3 (5C -3)(2n +2l +1)
\label{4.7t}
\ee
Of course, for both surface ($n=0$) and bulk ($n\neq 0$) modes these
frequencies differ 
from those of density fluctuations in the ideal one-component Fermi
gas.

	Damping of these modes will arise not only from scattering against
cold-atom impurities but also from thermal excitations (see Sect.~\ref{2.1st}. A
detailed discussion of the latter damping for the spin dipole excitation
has been given by Vichi and Stringari \cite{lorenzo}. 

\section{Summary and concluding remarks}
	In summary, the focus of this work has been on two-component
mixtures of fermionic atoms in dilute-vapour states and subject to
spherical harmonic confinement at zero temperature.
The main motivation has come from the experiments of DeMarco et al. [6] on
vapours of $^{40}$K atoms magnetically trapped in two different
hyperfine states. 
	
The generalized hydrodynamic equations of the mixture have allowed
us to discuss the damping mechanisms from correlations between partial
density fluctuations beyond mean field terms. Because of momentum
conservation in a pure two-component Fermi fluid the dissipation processes
in the hydrodynamic limit are associated with collisions between the two
components and these vanish quadratically with temperature because of Fermi
statistics. We have then discussed how a collisional regime may
nevertheless arise for both global and relative density fluctuations at
very low temperature from collisions of the Fermi fluid against cold
impurity atoms. We have seen that the establishment of a collisional
regime, in which the dynamical behaviour of the fluid reflects the quantal
statistics, is not subject to especially severe restrictions on the
strength of the fermion-impurity scattering nor on the number of impurities.

	We have then evaluated ground-state properties and small-amplitude
excitations of such a two-component Fermi fluid in a collisional regime. We
have shown that, whereas the role of the interactions in determining the
equilibrium density profiles is still very weak in the cases experimentally
studied so far, a rich phase diagram will emerge as the coupling strength
is increased and/or redistribution of the components between magnetic
states becomes allowed. The relevant coupling strength depends in a simple
manner on the number of fermions in the trap and on the ratio of the
scattering length to the harmonic-oscillator length. Finally, we have shown
how the problem of small-amplitude oscillations of both the total particle
density and the concentration density in a weakly coupled symmetric mixture
is amenable to full analytic solution in parallel with the analogous
problem for an ideal Fermi gas.

\small{This work is supported by the Istituto Nazionale di Fisica
della Materia 
through the Advanced Research Project on BEC. One of us (MA) wishes to
thank the Abdus Salam International Centre for Theoretical Physics for
their hospitality during the final stages of this work.}

\catcode `\@=11
\@addtoreset{equation}{section}
\def\theequation{\Alph{section}.\arabic{equation}}
\catcode `\@=12  
\appendix
\section{ Solution of equations~\refeq{4.3t} and \refeq{4.4t}}
\label{appdf}

	We give in this Appendix the details of the analytic solution of
the eigenvalue equations~\refeq{4.3t} and \refeq{4.4t}  and the
expressions for their 
eigenfunctions.

\subsection{Total density fluctuations}
\label{appdf1}

From Eqs.~\refeq{4.3t} and \refeq{3.8t} we find
\bea
\lefteqn{[6 (\omega/\omega_f)^2 +C \nabla_x^2(x^2)] \delta n(\vett x
,\omega)\nonumber} \\ &&+(3-C)(1-x^2) \nabla_x^2 \delta n(\vett x
,\omega)-(3-2C)\nabla(x^2)\cdot \nabla \delta n(\vett x
,\omega)=0
\label{a.1t}
\eea
where $x=r/R_F$. The solutions of Eq.~\refeq{a.1t} have the form
$\delta n ({\vett x}, \omega)=x^l F(x^2) Y_l^m(\theta, \phi)$, because
of spherical 
symmetry. Setting $x^2=y$, we determine the function $F(y)$ from
Eq.~\refeq{a.1t}  by means of
the Fuchs method for solving an ordinary differential equation in a series
form around regular singular points \cite{bender}. This method sets
\be
F(y)=(1-y)^s \sum_{k=0}^{\infty} a_k (1-y)^k
\label{a.2t}
\ee
and yields $s=C/(3-C)$ together with the recurrence relation for the
coefficients $a_k$,
\bea
\lefteqn{2(s+k+1)(s+k-3b+2)\frac{a_{k+1}}{a_{k}}=-3b(\omega/\omega_f)^2}
\nonumber
\\&& -3 (3b-1)+l (2-3b)+(s+k)[2(s+k-1)+2l-6b+7]\;.
\label{a.3t}
\eea
Here, $b=(3-C)^{-1}$. The eigenfunctions vanish at the boundary for
$C<3$.

	The eigenfrequencies are obtained from Eq.~\refeq{a.3t} by
asking that the 
solutions reduce to polynomials of degree $n+s$, i.e. $a_{n+1}=0$ for an integer $n$
representing the number of internal nodes of the density fluctuation
profile. This yields the dispersion relation reported in
Eq.~\refeq{4.6t} 
of the
main text.

\subsection{Concentration fluctuations}
\label{appdf2}

	From Eqs.~\refeq{4.4t} and~\refeq{3.8t} we get
\bea
\lefteqn{[6 (\omega/\omega_f)^2 +C \nabla_x^2(x^2)] \delta M(\vett x
,\omega)}\nonumber \\ &&+(5C-3)(1-x^2) \nabla_x^2 \delta M(\vett x
,\omega)+(3-4C)\nabla(x^2)\cdot \nabla \delta M(\vett x
,\omega)=0
\label{a.4t}
\eea
We look for solutions having the form $\delta M ({\vett x}, \omega)=x^l G(x^2) Y_l^m(\theta, \phi)$  and set $x^2=y$ to find the differential
equation obeyed by the function $G(y)$,
\bea
2(5C-3) y (1-y) \frac{d^2 G(y)}{dy^2} + \left[(5C-3)(2l+3)(1-y)+2 y(3-4C)\right] \frac{dG(y)}{dy} \nonumber \\ +\left(3 (\omega/\omega_f)^2+3 C + l (3-4C)\right) G(y)=0 
\label{a.5t}
\eea
Following again the Fuchs method we set
\be
G(y)=(1-y)^s \sum_{k=0}^{\infty} a_k (1-y)^k
\label{a.6t}
\ee
and from the indicial equation for Eq.~\refeq{a.5t} we find $s=C/(5C-3)$. Therefore, the
solutions will vanish at the boundary of the cloud only if $C>3/5$.

	The coefficients of the series in Eq.~\refeq{a.6t} obey the recurrence
relation
\bea
\lefteqn{2(s+k+1)[(s+k)(5C-3)-3+4C]
\frac{a_{k+1}}{a_{k}}=-3(\omega/\omega_f)^2}\nonumber
\\&& -[3C+l(3-4C)]+ (s+k)[(5C-3)(2s+2k+2l+1)+2(4C-3)]\;.
\eea
By asking again for polynomial solutions, we obtain the dispersion relation
for concentration fluctuations having $n$ internal nodes as given in Eq.~\refeq{4.7t} of the main text.


\end{document}